\documentclass[12pt]{article}
\usepackage{epsfig}

\begin{document}

\thispagestyle{empty}
\begin{center}
{\large \bf Effect of magnetic field on electron neutrino sphere in
pulsars } 
\vskip 3cm
{ \bf
Ashok Goyal \footnote{agoyal@ducos.ernet.in}}
\\

{ Department of Physics and Astrophysics\\
University of Delhi ,Delhi - 110 007  India \\}
{ Inter University Centre for Astronomy and Astrophysics\\
Ganeshkhind, Pune - 411 007  India \\}
\end{center}

\begin{abstract}
We study the neutrino interaction rates through charged as well as
neutral current weak interactions in hot, dense magnetized matter. At
densities near the neutrino sphere, matter in the presence of intense
magnetic field is polarized and electrons and protons typically occupy
the lowest Landau level. The weak interaction rates in such a
situation are severally modified leading to a sizable change in
neutrino opacity and consequently on the location of the neutrino
sphere which also gets distorted. This has implications for pulsar
velocity explanations. 
\end{abstract}

\pagebreak

%\begin{section}
Recently there has been a lot of interest \cite{vilenkin} - \cite{Nuno}
 in the study of neutrino  
transport in dense, hot nuclear matter in the presence of strong
magnetic  
fields reported to be present in young pulsars in connection with a
possible     mechanism to explain the recoil velocities of
pulsars. Specifically, an  
asymmetry arising due to parity violation in a strong magnetic field can 
lead to asymmetry in the explosion and recoil of newly formed neutron
stars.  
The asymmetry could for example, arise in the standard weak interactions 
of neutrinos \cite{vilenkin} - \cite{Hor}, due to neutrino magnetic moment
\cite{chugai}  or due to 
matter induced neutrino oscillations in the presence of magnetic field 
\cite{kusenko} - \cite{Nuno}. In order to explain the large velocities of
 neutron stars one 
needs an asymmetry of roughly one percent in the radiated neutrinos which  
could arise due to a number of neutrino scattering reactions on polarised  
electrons and hadron of the magnetized medium. A particularly attractive  
mechanism to explain pulsar kicks is suggested by Kusenko and Segre 
\cite{kusenko} and makes use of resonant MSW conversion.The way this 
mechanism works is as follows:
\par 
The high densities and temperatures reached in proto-neutron star
stage results in the trapping of neutrinos within their
neutrino-spheres i.e. the surface from which the optical depth of
neutrinos become one. The location of this neutrino-sphere is flavor
dependent because of the fact that electron neutrinos interact both
through charged as well as neutral currents, whereas the tau and mu
neutrinos have only the neutral current interactions. This results in
the tau neutrino escaping from deeper layers of star (due to their
large mean free path) where the temperature is high. The electron
neutrinos trapped between the tau and electron neutrino sphere can
then oscillate through resonant conversion to tau neutrinos and become
able to escape-thereby making the tau neutrino sphere to effectively
coincide with the resonance surface. The presence of magnetic field
modifies the potentials relevant for describing neutrino conversion
and leads to the distortion of resonance surface for MSW conversion
through polarisation of the magnetised medium. This leads to tau
neutrino emission in different directions from regions of different
temperatures, therefore with different energies and thus leading to
pulsar kicks. For this mechanism to work magnetic fields of strength 
$B \ge 10^{14}G$ are required and such high magnetic fields in the
context  
of supernova collapse have indeed been proposed in the literature
 \cite{Mueller}.

\par Such strong magnetic fields would affect the neutrino opacities by 
modifying the electron neutrino absorption and scattering
cross-sections  
on nucleons and leptons which determine the electron neutrino mean
free path. 
 The location of the $\nu_e$ sphere would thus get shifted and
distorted  
due to asymmetry in the cross-section. This effect on the dominant
absorption  
reaction $\nu_e + n \to p + e^-$ has been calculated in the literature 
\cite{Roulet} by modifying the phase space distribution of the final state 
electrons only and by leaving the matrix elements unchanged. In the 
presence of magnetic fields of the order of $10^{16} G$, the motion 
of the degenerate electrons at densities likely to be present near the 
neutrino sphere ($ Y_e \rho \sim 10^{10} - 10^{11} g/cc $)is in-fact
quantised  
and the electrons typically occupy the lowest Landau energy level 
leading to total polarisation of electrons opposite to the direction of 
the magnetic field. Because of charge neutrality viz $\bar{n}_e= n_p$ , 
the protons too are forced to occupy the lowest Landau level. In this 
situation , the matrix elements for the absorption and scattering
processes  
from nucleons and leptons get modified and have to be calculated by 
using the exact wave functions for electrons and protons by solving 
the Dirac equation in the magnetic field namely for the lowest Landau 
level $\nu=0$. Further in order to make numerical estimates of the 
neutrino mean free path in hot magnetised matter, we require to know 
the composition of nuclear matter. This is done by considering 
electrically neutral hot n-p-e matter in $\beta$- equilibrium in the 
presence of magnetic field. The nucleons here are treated 
non-relativistically and at the densities of interest, strong
interaction  
effects can be safely ignored.

\par  The processes that contribute to $\nu_e$ opacity are the
absorption  
\begin{equation}
\label{one}
\nu(p_1)+n(p_2) \rightarrow p(p_3)+ e^-(p_4)
\end{equation}
and the scattering
\begin{eqnarray}
\label{two}
 \nu(p_1)+n(p_2) & \rightarrow & n(p_3)+ \nu(p_4)   \\
\label{three}
 \nu(p_1)+p(p_2) & \rightarrow & p(p_3)+ \nu(p_4)   \\
\label{four}
 \nu(p_1)+e^-(p_2) & \rightarrow & e^-(p_3)+ \nu(p_4)  
\end{eqnarray}

processes.
\par  The cross-section per unit volume of matter or the inverse mean
free  
path is given by
\begin{equation}
\label{five}
\frac{\sigma(E_1)}{V} = \lambda^{-1}(E_1) 
= \frac{1}{2E_1} \Pi_{n=2,3,4} d \rho_n  W_{fi} f_{2}(E_2)
(1-f_{3}(E_3))(1-f_{4}(E_4))
\end{equation}
where $d \rho_n = \frac{d^3p_n}{(2\pi)^3 2E_n}$ is the density of
states of  
particles with four momenta $p_n, f_n(E_n)$ are the particle
distribution  
functions which in thermal equilibrium are given by the usual
Fermi-Dirac  
distributions $f_n(E_n) = 1/(1+exp(\frac{E_n - \mu_n}{T}))$ and 
($1-f_n(E_n)$) accounts for the Pauli Blocking factor for the final
state  
particles. 
The transition rate $W_{fi}$ is
\begin{equation}
\label{six}
W_{fi} = (2\pi)^4\delta^4(P_f-P_i)\left\vert M \right\vert^2
\end{equation}
where $\left\vert M\right\vert^2$ is the squared matrix element summed over
 initial and
final spins. 
In the presence of magnetic field, the density of states are replaced by 
\begin{equation}
\label{seven}
d\rho_n =\sum_{\nu} (2-\delta_{\nu,0})\int_{-\infty}^{\infty}\int_
{-\frac{eBL_X}{2}}^{\frac{eBL_X}{2}} \frac{dp_{nz}dp_{ny}}{(2\pi)^2 2E_n}
\end{equation}
where the sum over the Landau levels is to be performed. If the magnetic 
field is not strong enough to force particles in the lowest $\nu=0$
Landau  
state, the matrix element remains essentially unchanged \cite{Matese} and one
has  
to only take the proper phase space into consideration. In the
presence of  
quantising magnetic field, a situation likely to be present for strong
 fields at densities and temperatures under consideration, the electrons 
occupy the lowest Landau level and the matrix elements have to be
calculated  
by using the exact wave functions of relativistic electrons and
protons in  
a magnetic field.
In a gauge in which the vector potential $A =(0,xB,0)$, the quantum
states are  
specified by the quantum numbers $p_y,p_z,\nu and s $ and the energy is
given  by \cite{Itz}
\begin{equation}
\label{eight}
E = \sqrt{m^2+p_z^2+2\nu eB}
\end{equation}
and the positive energy electron spinor in $\nu = 0 $ state is given by
\begin{equation}
\label{nine}
u_e = \left(
\begin{array}{c}
0 \\
E_e+m_e \\
0 \\
-p_{eZ} \\
\end{array}
\right)
\end{equation}
\vskip 0.3cm
electrons being alligned antiparallel to the direction of the magnetic
field  
which is along the Z axis. The x-component of the momentum is not
conserved 
 now and the transition rate becomes
\begin{equation}
\label{ten}
W_{fi} =
(2\pi)^3\delta(E_f-E_i)\delta(P_{fz}-P_{iz})\delta(P_{fy}-P_{iy})\left\vert M 
\right\vert^2
\end{equation}
$V = L_xL_yL_Z$ is the normalization volume.
Squared matrix element summed over initial and final spins for processes
(1)-(4) in the Standard model is given by
\begin{equation}
\label{eleven}
\left \vert M \right\vert^2 = 32G_F^2\Bigg[ (C_V+C_A)^2 p_1.p_2p_3.p_4 +
(C_V-C_A)^2p_2.p_4p_1.p_3 - m_2m_3(C_V^2-C_A^2)p_1.p_4\Bigg] 
\end{equation}
the vector and the axial vector couplings are given by
\[  {\begin{array}{ccc}
\hline
Process & C_V & C_A    \\
\hline        \\
\nu n \to p e    & g_V=1              & g_A \approx 1.22    \\
 \nu n \to n \nu & -g_V/2             & -g_A/2  \\
 \nu p \to p \nu & 1/2-2sin^2\theta_W \approx 0 & g_A/2  \\
 \nu e \to e \nu & 1/2+2sin^2\theta_W  & 1/2  \\
\hline
\end{array}}\]
\vskip.3cm

\par  During the Kelvin-Helmholtz cooling phase of the proto-neutron
star, the density is much less than the nuclear density. The nucleons
are non-relativistic and non-degenerate, the electrons remain
relativistic. The neutrinos are non-degenerate and the elastic
approximation in the scattering processes is quite reliable. The phase
space integrals in eqn(5) can then be carried out analytically \cite 
{Tubb}
and we obtain the mean free paths as  
\begin{eqnarray}
\label{twelve}
\lambda_{abs}^{-1}(0) &=& \frac{G_F^2}{\pi}(g_V^2+3g_A^2)E_\nu^2n_N\frac{1}
{1+exp(E_\nu+Q-\mu_e)\beta} \\
\lambda_n^{-1}(0) &=& \frac{G_F^2}{4\pi}(g_V^2+3g_A^2)E_\nu^2n_N  \\
\lambda_p^{-1}(0) &=& \frac{3G_F^2}{4\pi}g_A^2E_\nu^2n_p  \\
\lambda_e^{-1}(0) &=& \frac{2G_F^2}{3\pi^3}(C_V^2+C_A^2)\mu_e^2TE_\nu^2  
\end{eqnarray}
\par  Considering now the magnetic field that affects only the electrons
 and which is not strong enough to confine the electrons to the lowest Landau
 state, the matrix elements for the processes remain essentially unchanged
 and modifying only the phase space, the important absorption cross-section
 turns out to be \cite{Roulet}
\begin{eqnarray}
\frac{\sigma_{abs}(B)}{V}
& \sim & 
 \frac{G_F^2}{2\pi}(g_V^2+3g_A^2) eB \frac{1}{1+exp(E_\nu+Q-\mu_e)\beta}
\nonumber \\
&&   (E_\nu+Q)\sum (2-\delta_{\nu,0})\frac{1}{\sqrt{(E_\nu+Q)^2-m_e^2-2\nu eB}}
\end{eqnarray}
\vskip.3cm
The cross-section for electron neutrino scattering in hot dense matter in
 the presence of magnetic field has been derived in \cite{bez1}. The other
 processes remain unaffected by the magnetic field.
\par  In the interesting case of strongly quantised magnetic field when both 
electrons and protons in stellar matter are polarised with their spins aligned
anti parallel and parallel to the direction of the magnetic field respectively,
 the matrix elements for the above processes are calculated using exact wave
functions for $\nu = 0$ Landau state .For non-relativistic nucleons and
 relativistic leptons, we get for the matrix element squared
\begin{eqnarray}
M^2_{abs} &=& 8G_F^2m_2m_3(E_4+p_{4z})\Bigg[{(g_V+g_A)^2+4g_A^2}E_1 +
 {(g_V-g_A)^2+4g_A^2}E_1Cos\theta\Bigg]           \nonumber \\
&&  exp\Bigg[-\frac{1}{2eB}{(p_{1x}+p_{2x})^2+(p_{3y}+p_{4y})^2}\Bigg]
                                        \\
M^2_{p\nu}                                
&=& 16G_F^2{m_2}^2(p_1.p_4+2p_{1Z}.p_{4Z}) \nonumber \\
&&  exp\Bigg[-\frac{1}{2eB}{(p_{4x}-p_{1x})^2+(p_{4y}-p_{1y})^2}\Bigg]
       \\
M^2_{e\nu} 
&=& 16G_F^2\Bigg[(C_V^2+C_A^2)\Bigg[(E_1E_4 + p_{1Z}p_{2Z})
(E_2E_3+p_{2Z}p_{4Z})       \nonumber \\
&& -(E_1p_{4Z}+E_4p_{1Z})(E_2p_{3Z}+E_3p_{2Z})\Bigg]
           \nonumber \\
&&  +2C_VC_A \Bigg[(E_1E_4+p_{1Z}p_{4Z})(E_2p_{3Z}+E_3p_{2Z}) \nonumber \\
&& -(E_1p_{4Z}+E_4p_
{1Z})(E_2E_3+p_{2Z}p_{3Z})\Bigg]\Bigg]  \nonumber \\
&&  exp\Bigg[-\frac{1}{2eB}{(p_{4x}-p_{1x})^2+(p_{4y}-p_{1y})^2}\Bigg]
\end{eqnarray}
\vskip.3cm
The scattering cross-sections are now given by
\begin{eqnarray}
\frac{\sigma_{scatt}(B)}{V} &=& \lambda_{scatt}^{-1}  \nonumber \\ 
 &=&
\frac{1}{2E_1L_X}\int\frac{d^3p_4}{(2\pi)^32E_4}
\int \frac{dp_{2Y}dp_{2Z}}{(2\pi)^22E_2}
 \int\frac{dp_{3Y}dp_{3Z}}{(2\pi)^22E_3} \nonumber \\
&&  (2\pi)^3\delta(P_Y)\delta(P_Z)\delta(E) |M|^2  \nonumber \\
&& f_2(E_2)(1-f_3(E_3))(1-f_4(E_4))            
\end{eqnarray}
\vskip.3cm
and the absorption cross-section is obtained by exchanging the fourth and the
 second particles. The integrals over y-component of the momenta can be done
 by using the y-component momentum conserving delta function. The rest of the
 integrals can be done by using $\delta(E)$ and $\delta(P_Z)$ in the limit of 
non-degenerate, non-relativistic nucleons. The exponential functions in the 
squared matrix elements can be approximated by one in this limit and further
 in the limit of elastic scattering approximation. Carrying out the integrals
 we obtain
\begin{eqnarray}
\frac{\sigma_{abs}(B)}{V}
 &=& \lambda_{abs}^{-1}(B)
 \sim  
 \frac{G_F^2}{4\pi}eB n_N\Bigg[\{(g_V+g_A)^2+4g_A^2\}+\{(g_V+g_A)^2-4g_A^2\}
cos\theta\Bigg]
            \nonumber \\
&&  \frac{1}{1+exp\{-(E_\nu+Q-\mu_e)\beta \}}
\end{eqnarray}
\begin{eqnarray}
\frac{\sigma_{\nu p}(B)}{V}
&=&  \frac{2G_F^2C_A^2eB}{(2\pi)^3}\int_{0}^{\infty} dp_{2Z}f_2(E_2)
(E_\nu+\frac{p_{2Z}^2}{2m_p})\Bigg[(E_\nu +\frac{p_{2Z}^2}{2m_p})
 \nonumber \\
&& \{2-\frac{1}{3m_p}(E_\nu+\frac{p_{2Z}^2}{2m_p})\}
-\frac{p_{nuZ}^2+p_{2Z}^2}{m_p}\Bigg]
\end{eqnarray}
which can be approximated by 
\begin{equation}
\frac{\sigma_{\nu p}(B)}{V} \approx \frac{G_F^2C_A^2}{\pi}n_pE_\nu^2
\end{equation}
where      
\begin{equation}
n_p = \frac{2eB}{(2\pi)^2}\int_{0}^{\infty} f_2(E_2)dp_{2Z}
\end{equation}
\begin{eqnarray}
\frac{\sigma_{\nu e}(B)}{V}
 & \sim & 
\frac{4G_F^2eB}{(2\pi)^3}\mu_eT\Bigg[(C_V^2+C_A^2)(1+cos^2\theta)-4C_VC_Acos
\theta\Bigg]         \nonumber \\
&& \frac{E_\nu}{1+exp(-\mu_e/T)}
\end{eqnarray}
\vskip.3cm
The neutrino-neutron scattering cross-section remains unchanged except
 through its dependence on neutron density which gets affected by the
 magnetic field.
\par  For the noninteracting, hot n-p-e matter in beta-equilibrium we have
 the number densities given by
\begin{eqnarray}
\bar{n}_e = n_{e^-}-n_{e^+} 
&=&  \frac{2eB}{(2\pi)^2}\Bigg[\sum_{\nu}(2-\delta_{\nu,o})\int_{0}^
{\infty}dp_Z \frac{1}{1+exp\{\beta(E_{e\nu}-\mu_e)\}}  \nonumber \\
&& -(\mu_e \to -\mu_e)\Bigg]
\\
n_p
&=&  \frac{2eB}{(2\pi)^2}          \nonumber \\
&& \Bigg[\sum_{\nu}(2-\delta_{\nu,o})
\int_{0}^{\infty} dp_Z\frac{1}{1+exp\{\beta (E_{p\nu}-\mu_p)\}}\Bigg] \\   
n_n                               
&=&  \frac{1}{\pi^2}\int_{0}^{\infty}p^2dp\frac{1}
{1+exp\{\beta(E_{n\nu}-\mu_n)\}}   
\end{eqnarray}
\vskip.3cm
The energy expressions in the presence of magnetic field are given in eqn(9).
 Charge neutrality requires $\bar{n}_e=n_p$ and $\beta$-equilibrium relates the
 chemical potentials viz. $\mu_n=\mu_p+\mu_e$
\par  In Fig.1 we have plotted the proton fraction $Y_p=\frac{n_p}{n_n+n_p}$
as a function of density at two different temperatures 1 MeV and 10
 MeV
 respectively for different values of the magnetic field. The neutron
 fraction is given by $Y_n=1-Y_p$. At T=1 MeV, there are very small number 
of positrons present and the effect of the magnetic field is to raise the
 proton fraction so much so that at $B=10^4 MeV$, the matter becomes
 predominantly proton rich. This had the effect of not only modifying 
neutrino cross-sections quantitatively but to change the relative importance
 of various processes in the presence of magnetic field. In Fig.2 we have
 plotted the neutrino mean free path in meters as a function of baryon density
 for different values of the magnetic fields for matter at T=1 MeV. The 
neutrino energy is taken to be $E_1=3T$. We observe as expected that neutrino
 absorption cross section dominates the mean free path even in the presence
 of highest magnetic fields considered here as it does in the free field case.
 We also find that the absorption mean free path rises rapidly with density
 since at low temperatures due to Pauli-Blocking effect, the absorption 
process gets suppressed. At higher temperatures, there are lots of positrons 
present and electron degeneracy decreases and the effect is absent as can be 
seen from Figures 3 and 4 where we have plotted the mean free paths for matter
 at T=10 and T=30 MeV respectively. In fact at these temperatures, the free 
path decreases with increase in density. The absorption and neutrino 
electron scattering cross sections develop asymmetry when the matter is 
polarised as can be seen from eqns.(21) and(25). The asymmetry in the 
absorption cross section for neutrinos propagating along and opposite to 
the direction of the magnetic field is roughly of the order of 20 \%. 
The asymmetry is far more pronounced in neutrino electron scattering , 
so much so that scattering cross section for neutrinos in the direction 
of the magnetic field is ten times more compared to neutrinos progating 
parallel to the field. Thus at low temperatures, when the matter is highly 
polarised the neutrino sphere itself gets severely modified and distorted. 
This results in neutrinos  along the magnetic field being emitted from
deeper layers where the temperature is higher in comparison to
neutrinos which are emitted anti parallel to the field direction. A
change in cross sections ( by an order of magnitude even at high
temperatures ) caused by a large magnetic field will change the
density near the neutrino sphere considerably. We can see from Figures
2, 3 and 4 that increase in neutrino cross section is roughly
proportional to B ( albeit for polarised matter ) thereby shifting the
neutrino sphere to larger radii and hence to lower densities. This
will have implications for the implementation of Kusenko-Segre
mechanism to explain the observed pulsar velocities.
\vskip 0.5cm
{\bf Acknowledgement} \\
I am thankful to the organisers of WHEPP-5 (Fifth workshop on High
Energy Physics Phenomenology held from JAN.12-20 1998 at
Inter-University Centre for Astronomy and Astrophysics, Pune, India)
where this problem was conceived and discussed in the working group.

\pagebreak

\pagebreak
{\bf Figure Captions} \\
\vskip.5cm
Fig.1  Proton fraction $Y_p$ as a function of baryon density in gm/cc 
       for matter
       at $T=1$ MeV and $T=10$ MeV respectively. The solid, long dashed, 
      dashed 
       and dotted curves are for $B=0,10^2,10^3$ and $ 10^4 MeV^2$ respectively.
\vskip.2cm
Fig.2  Neutrino mean free path in meters as a function of baryon density for 
       matter at T=1 MeV and $B=0,10^2$ and $10^4 MeV^2$. The neutrino energy 
       is taken to be 3T. The solid curve is for the absorption process and
       the long dashed, dashed and dotted curves are for neutrino scattering
       on neutrons, protons and electrons respectively.
\vskip.2cm
Fig.3  Neutrino mean free path in meters as a function of baryon density for 
       matter at T=10 MeV and $B=0$ and $10^4 MeV^2$. The neutrino energy 
       is taken to be 3T. The solid, long dashed, dashed and dotted curves are
       as in Figure2.
\vskip.2cm
Fig.4  Neutrino mean free path in meters as a function of baryon density for 
       matter at T=30 MeV and $B=0$ and $10^4 MeV^2$. The neutrino energy 
       is taken to be 3T. The solid curve, long dashed, dashed and dotted
       curves are as in Figure2.

%
% inclusion of figures in paper
%  dated : 16-12-1998
%
%\begin{center}

\begin{figure}
\epsfig{file=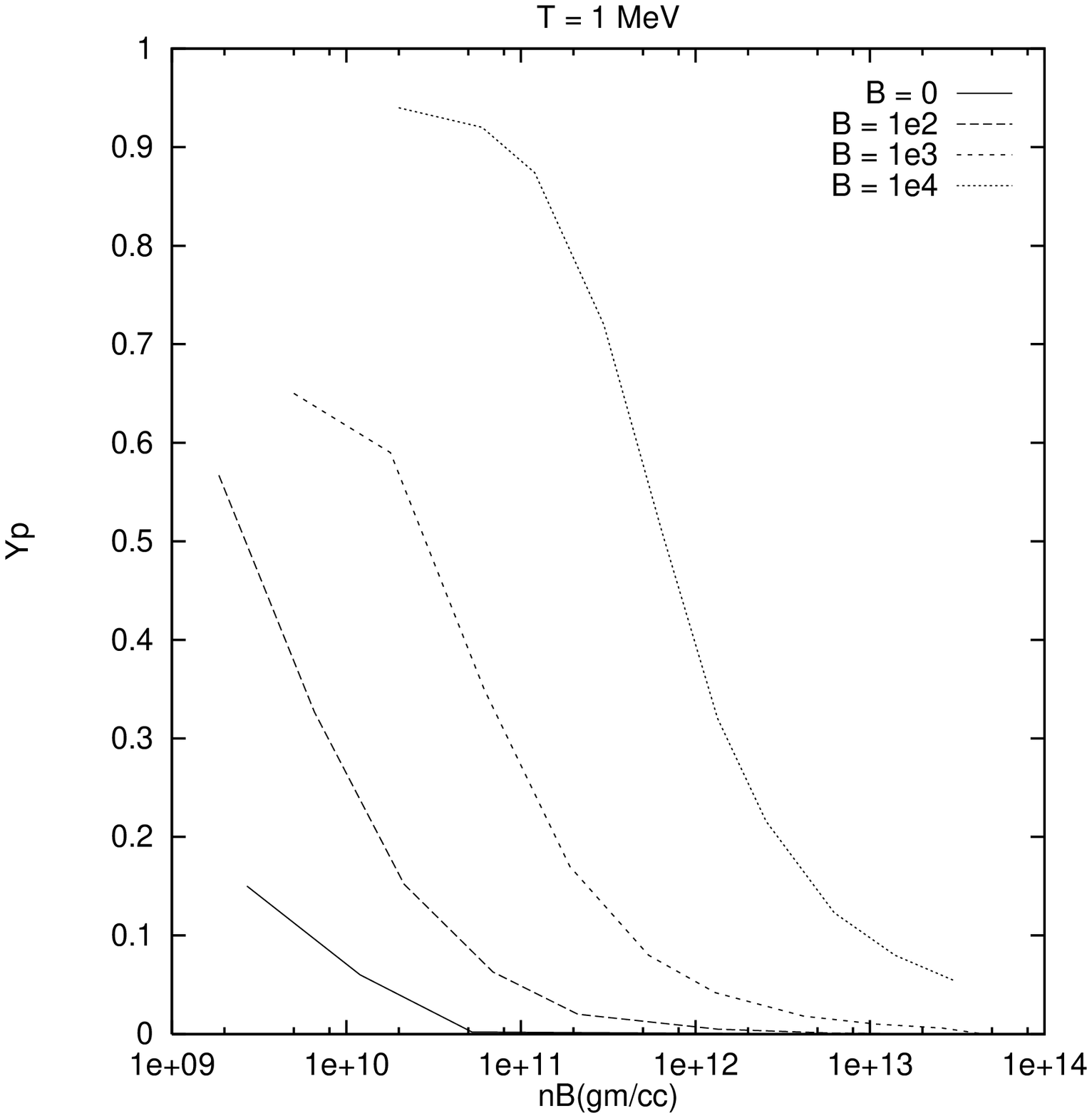,width=10cm,height=8cm}
\vskip 0.3cm
\epsfig{file=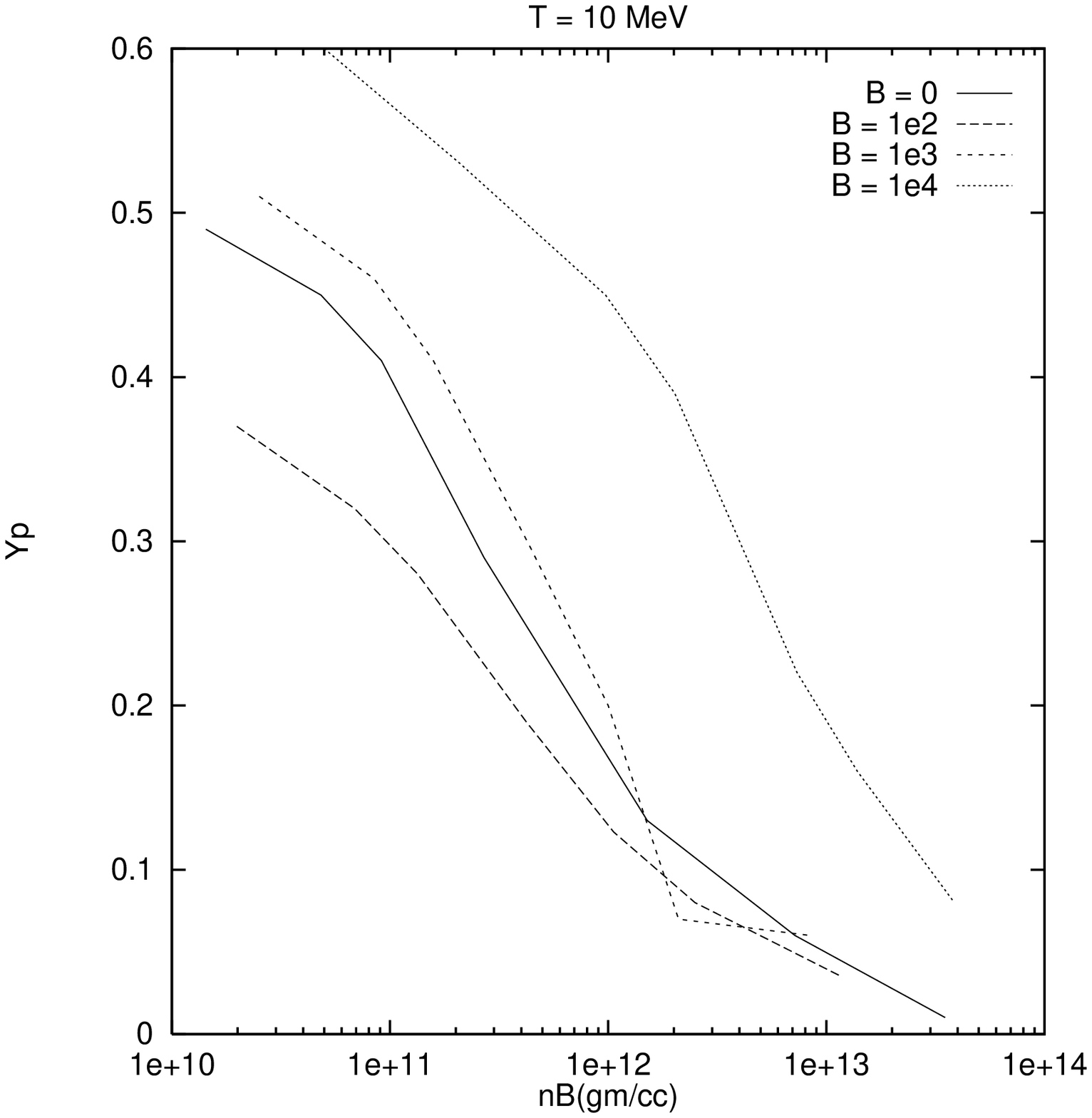,width=10cm,height=8cm}
\caption{}
\end{figure}
\begin{figure}
\epsfig{file=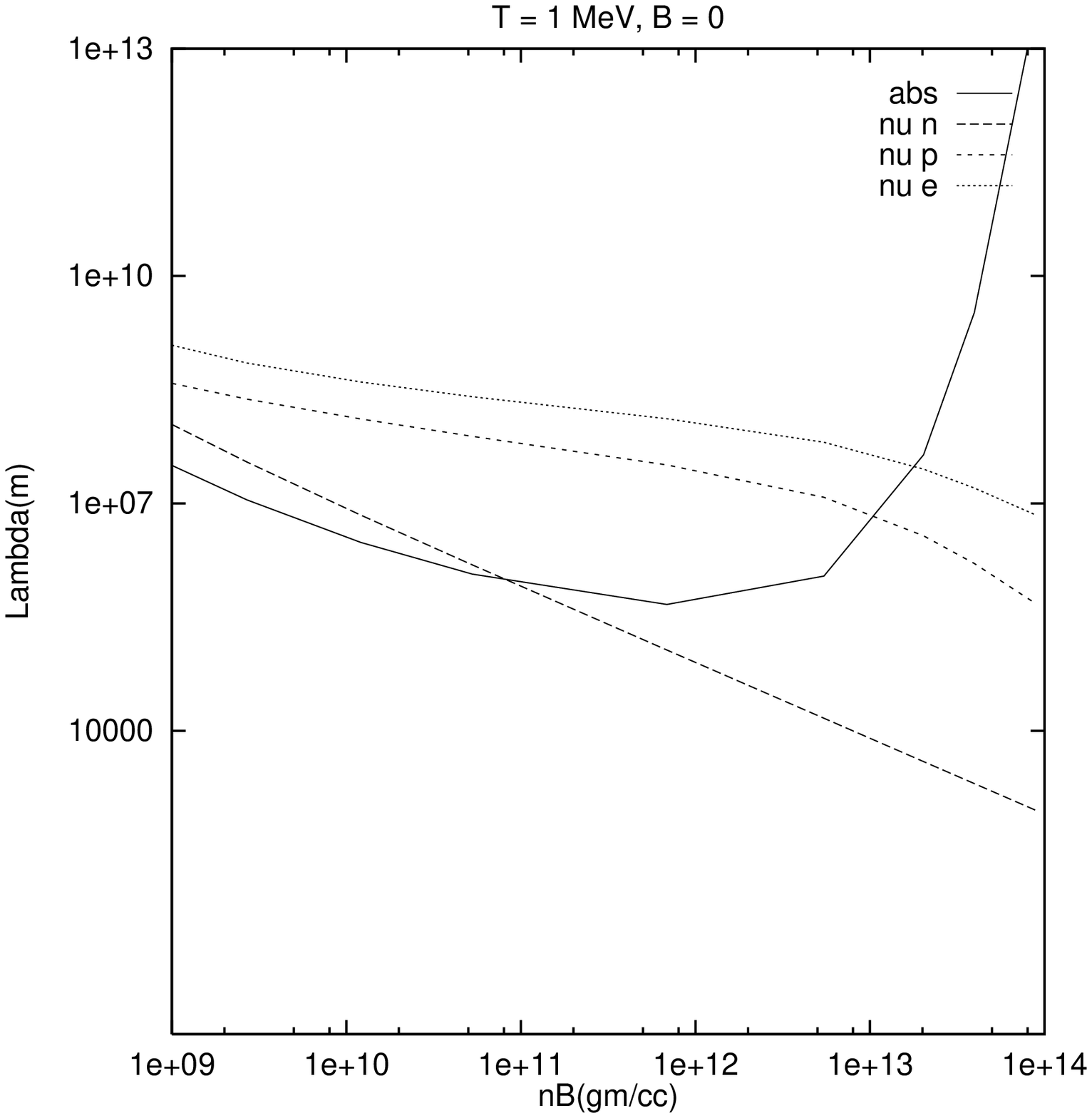,width=10cm,height=8cm}
\vskip 0.3cm
\epsfig{file=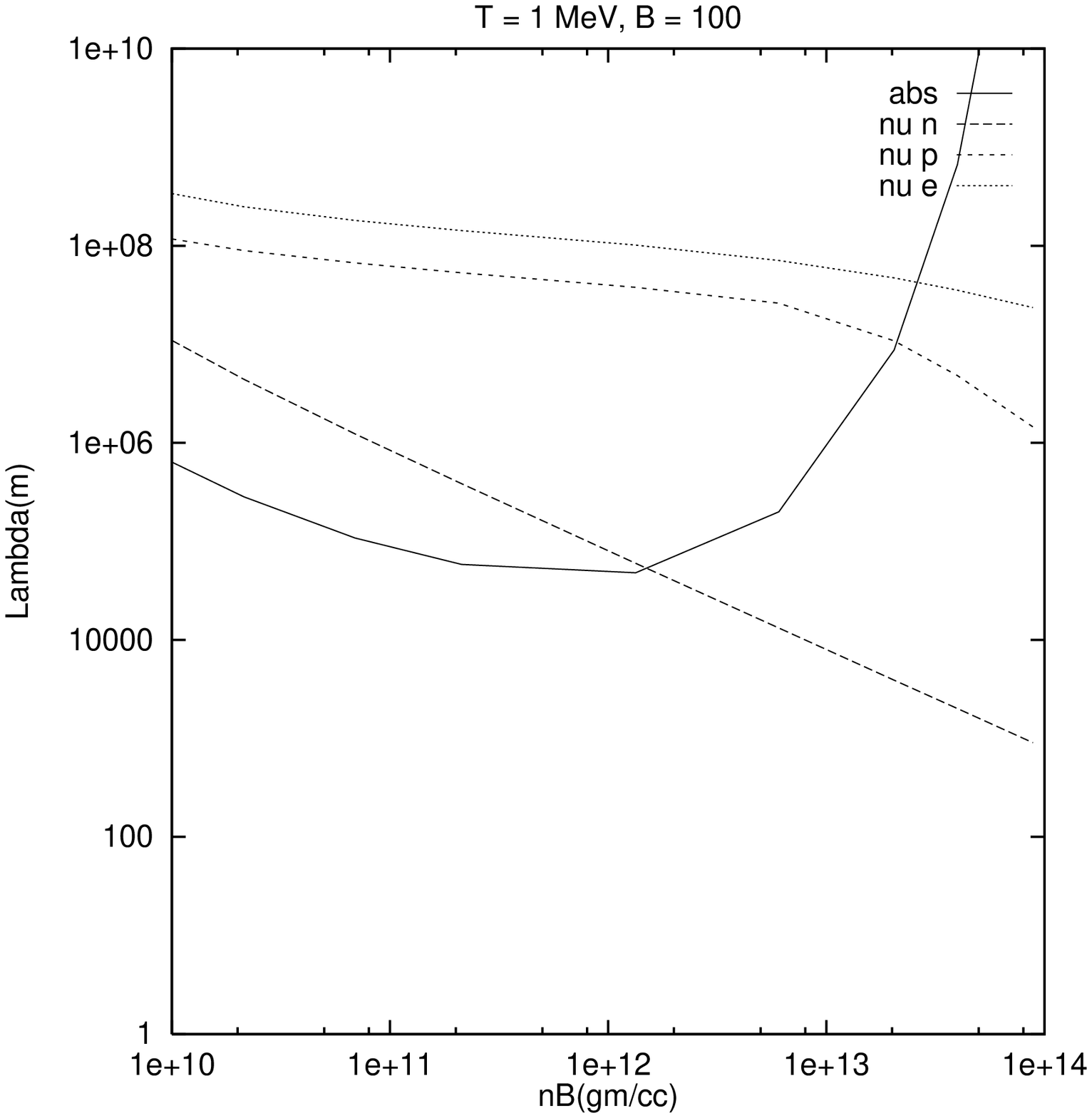,width=10cm,height=8cm}
\vskip 0.3cm
\epsfig{file=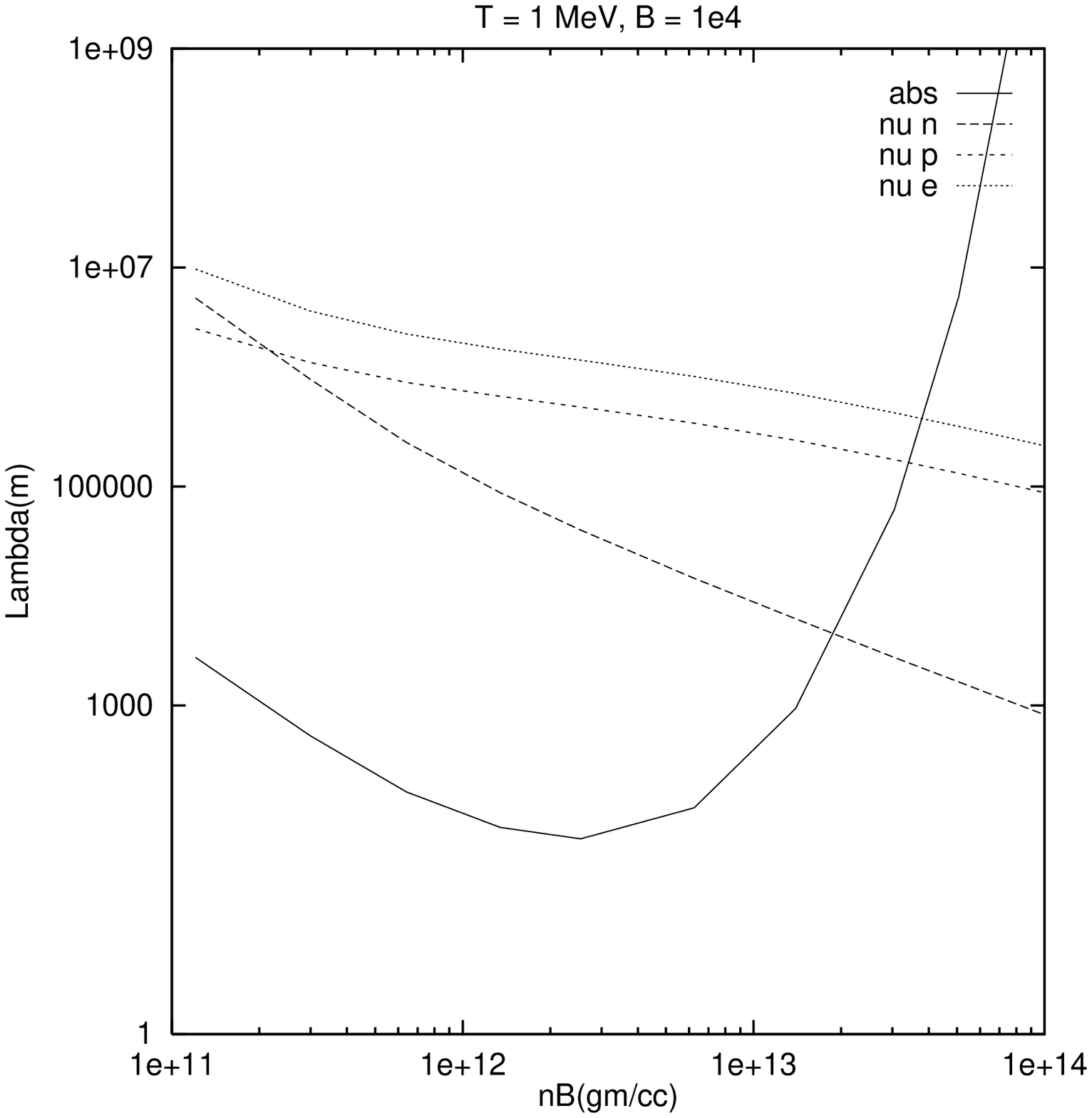,width=10cm,height=8cm}
\caption{}
\end{figure}
\begin{figure}
\epsfig{file=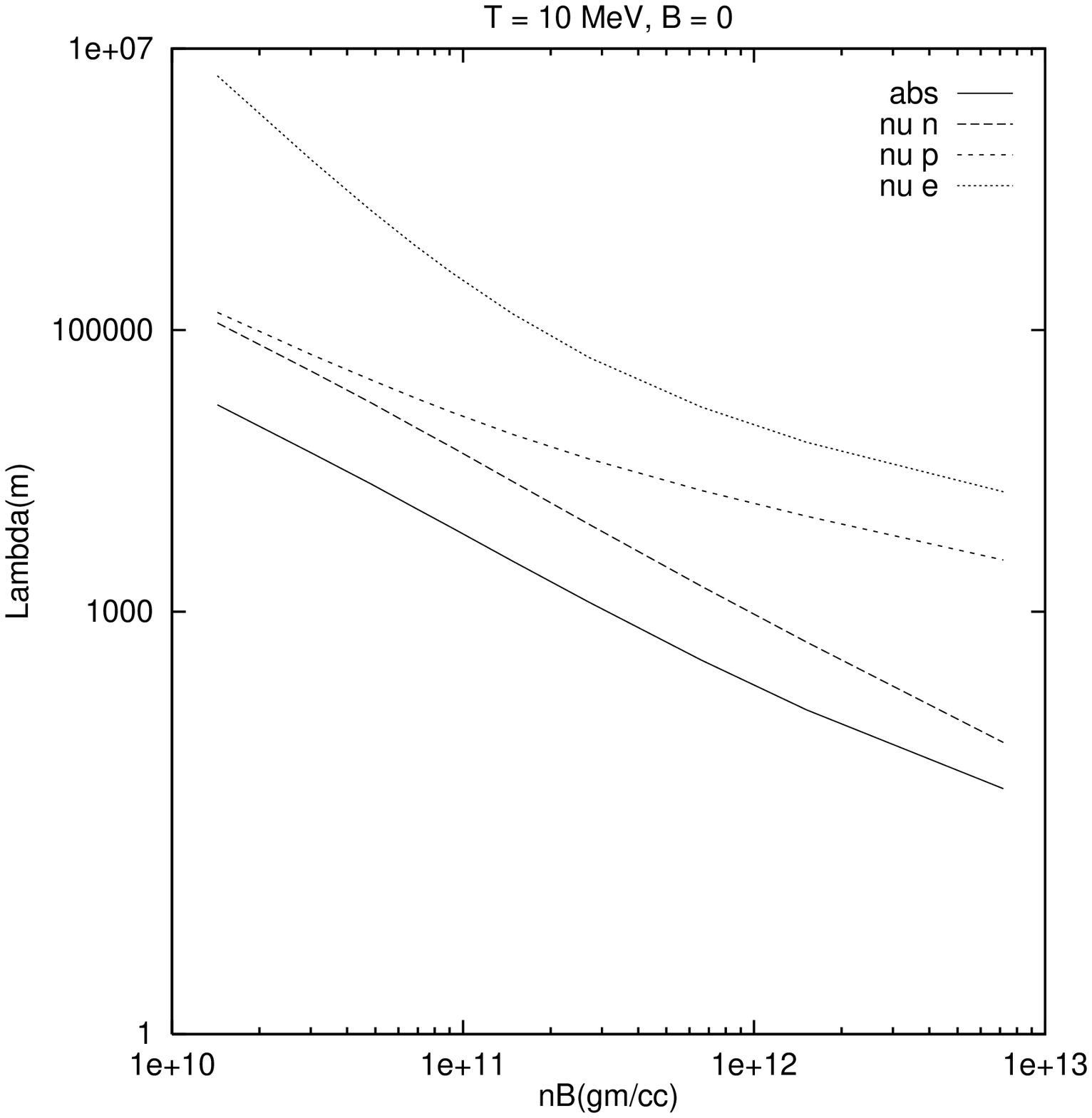,width=10cm,height=8cm}
\vskip 0.3cm
\epsfig{file=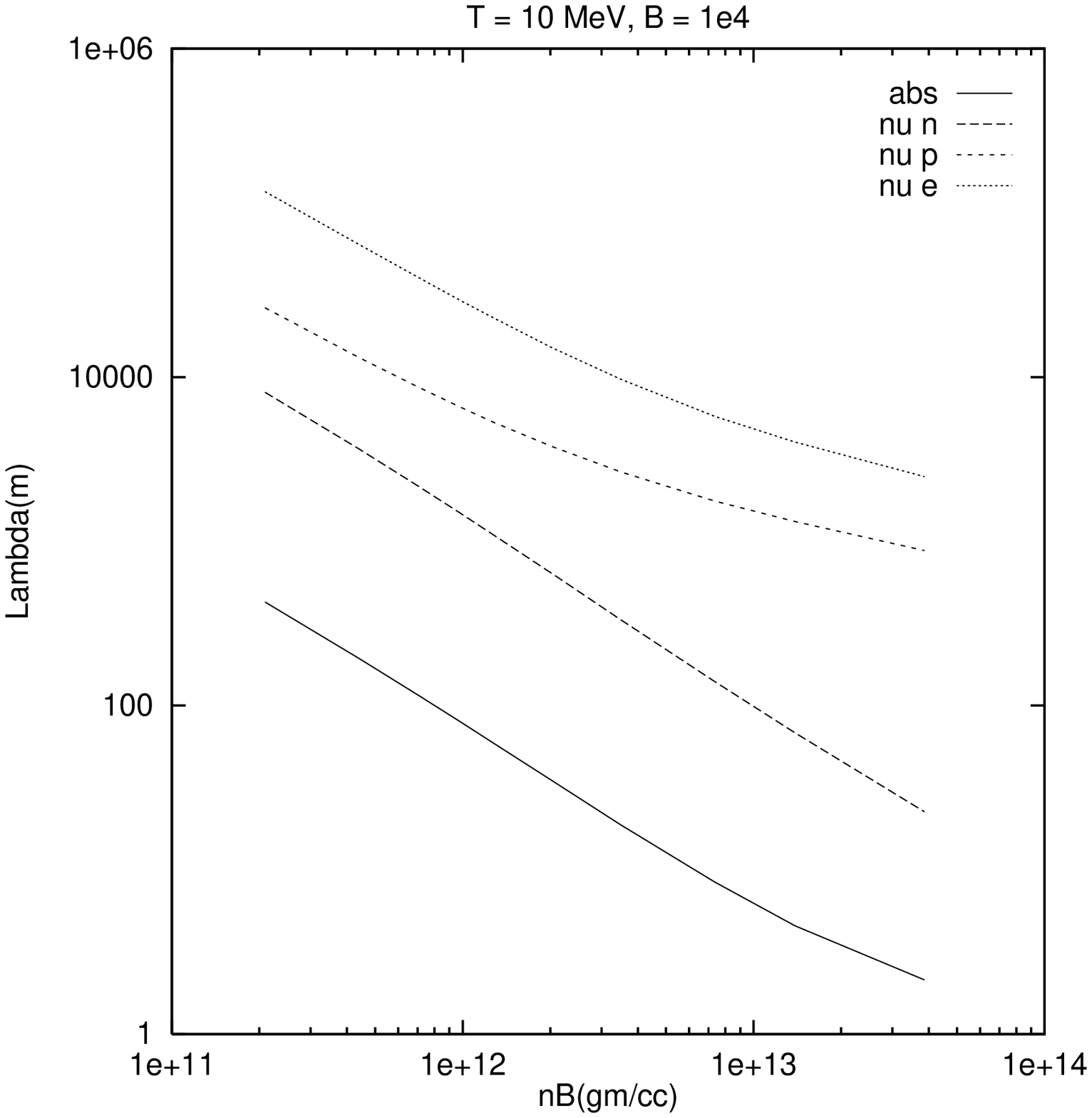,width=10cm,height=8cm}
\caption{}
\end{figure}
\begin{figure}
\epsfig{file=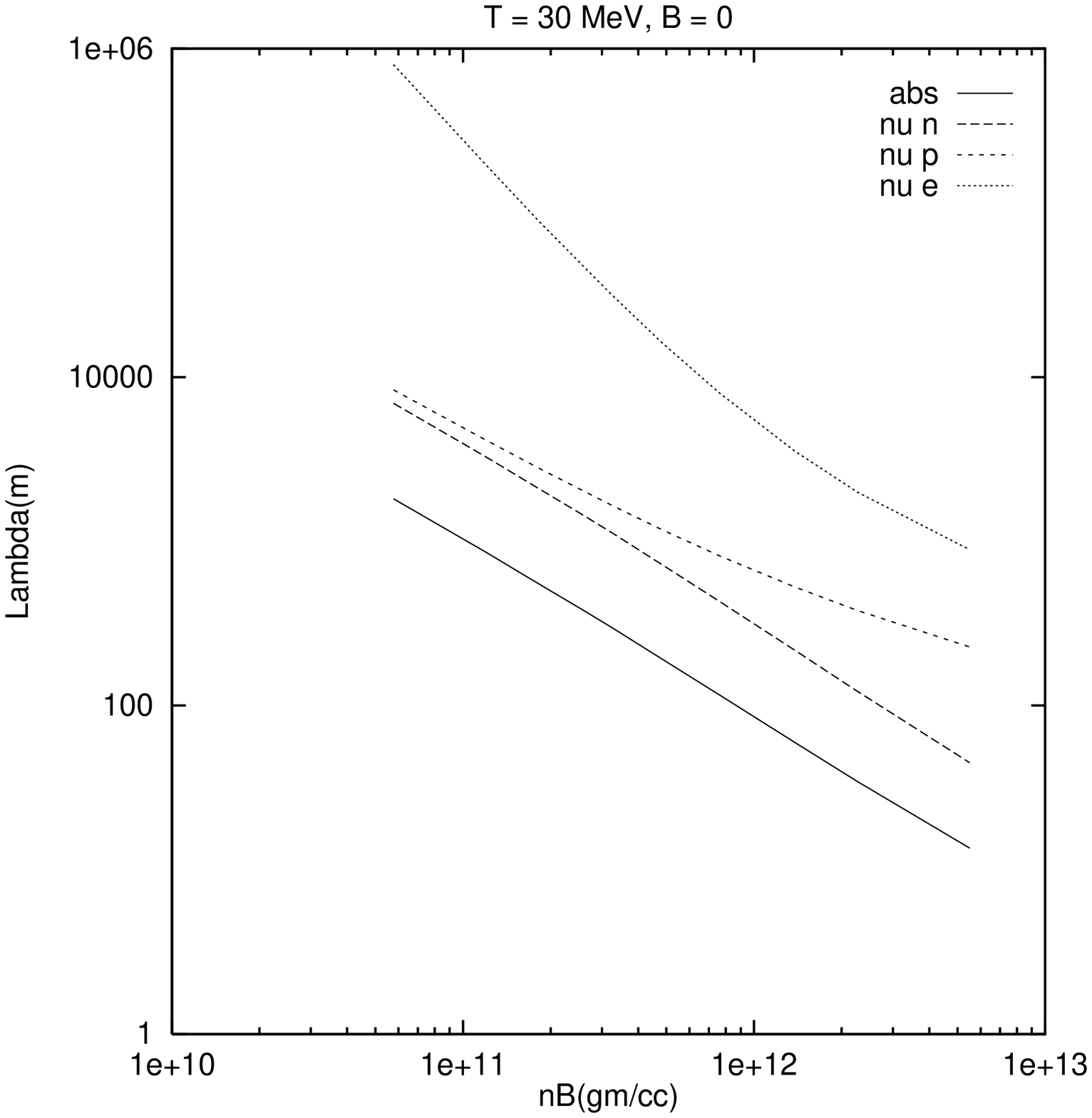,width=10cm,height=8cm}
\vskip 0.3cm
\epsfig{file=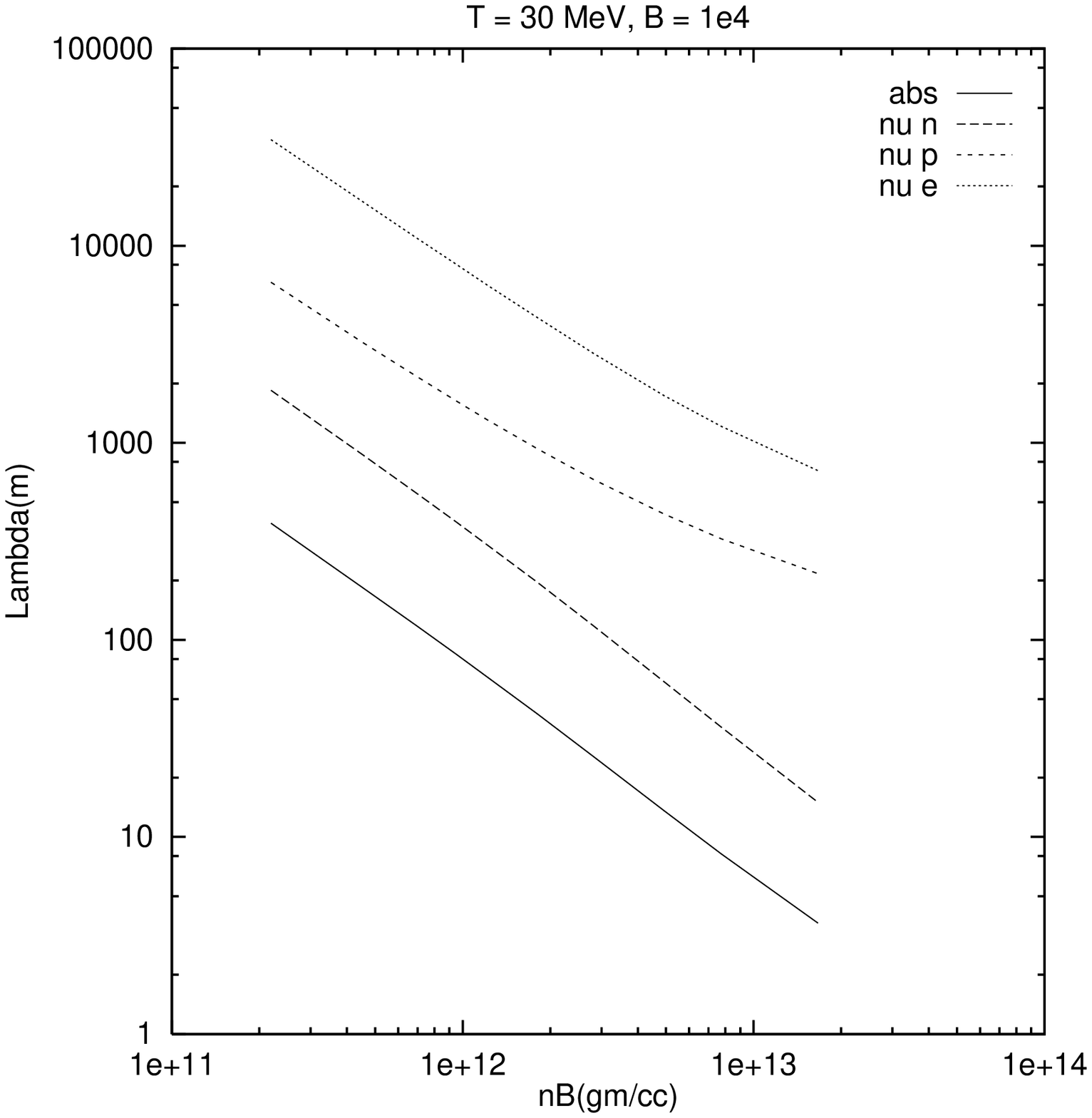,width=10cm,height=8cm}
\caption{}
\end{figure}
%\end{center}
%EndOfFile


\begin{thebibliography}{plain}
\bibitem{vilenkin} 
A.Vilenkin,{\em Astroph.J. }{\bf 451}(1995)700
   A.V.Kuznestov and N.V.Mikhev, hep-ph/9612312
   G.S.Bisnovatyi-Kogan, hep-ph/9707120

\bibitem{bez1} 
  V.G.Bezchastnov and P.Haensel, {\em Phys.Rev.}{\bf D54}(1996)3706

\bibitem{Hor}
 C.J.Horowitz and J.Piekarewicz, hep-ph/9701214

\bibitem{kusenko}
 A.Kusenko and Gino Segre,{\em Phys.Rev.Lett.} {\bf 77}, 4872 (1996)

\bibitem{Nuno}
H.Nunokawa,V.B.Semikoz,A.Yu.Smirnov and J.W.F.Valle, {\em Nucl.Phys.}
 {\bf B501} , 17(1997)

\bibitem{chugai}
N.N.Chugai,{\em Sov.Astron.Lett.}{\bf 10}, 87 (1984)

\bibitem{Mueller}
E.Mueller and W.Hillebrandt,{\em Astrn.and Astroph.} {\bf 80}, 147 (1979)
C.Thompson and R.C.Duncan,{\em Astrophy.J.}{\bf 392}, 19 (1992)
G.S.Bisnovatyi and S.G.Moiseenko,{\em Sov.Astron.}{\bf 36}, 285 (1992)
M.Bocquet et.al.,{\em Astron. and Astrophy.}{\bf 301}, 757 (1995)

\bibitem{Roulet}
Esteban Roulet, hep-ph/9711206

\bibitem{Matese}
J.J.Matese and R.F.Connell,{\em Phys.Rev.}{\bf 180}, 1289 (1969)
L.Fassio-Canuto,{\em Phys.Rev.}{\bf 187}, 2141 (1969)
D.Lai and S.L.Shapiro,{\em Astroph.J.}{\bf 383}, 745 (1991)

\bibitem{Itz}
C.Itzykson and J-B.Zuber,{\em Quantum Field Theory}
(McGraw Hill,New York,1985) p.67

\bibitem{Tubb}
See for example
D.L.Tubbs and D.N.Schramm,{\em Astrophys.J.}{\bf 201}, 467 (1975)
A.Burrows and J.M.Lattimer,{\em Astrophys.J.}{\bf 307}, 178 (1986)

\end{thebibliography}
\end{document}